\documentclass[conference]{IEEEtran}
\usepackage{amsmath,amssymb,amsfonts}
\usepackage{booktabs}
\usepackage{graphicx}
\usepackage{array}
\usepackage[svgnames]{xcolor}
\usepackage{tikz}
\usepackage{tabularx}
\usetikzlibrary{arrows.meta,calc,positioning,backgrounds}
\usepackage[hidelinks]{hyperref}
\usepackage{cite}
\usepackage{pgfplots}\pgfplotsset{compat=1.17}
\usepackage{algorithm}
\usepackage{algpseudocode}
\definecolor{cblue}{RGB}{41,128,185}\definecolor{cred}{RGB}{192,57,43}
\definecolor{corange}{RGB}{211,84,0}\definecolor{cgray}{RGB}{127,140,141}

\usepackage[normalem]{ulem}
\newcommand{\delm}[1]{}
\definecolor{revadd}{RGB}{0,90,200}
\definecolor{revdel}{RGB}{200,30,30}
\definecolor{revflag}{RGB}{150,45,160}
\newif\ifrevfinal
\revfinaltrue   
\ifrevfinal
  \definecolor{revadd}{RGB}{0,0,0}
  \definecolor{revdel}{RGB}{0,0,0}
  \definecolor{revflag}{RGB}{0,0,0}
  \newcommand{\ra}[1]{#1}        
  \newcommand{\rd}[1]{}          
  \newcommand{\rf}[1]{}          
\else
  \newcommand{\ra}[1]{\textcolor{revadd}{#1}}       
  \newcommand{\rd}[1]{\textcolor{revdel}{#1}}
  \newcommand{\rf}[1]{\textcolor{revflag}{\textsuperscript{[}\,#1\,\textsuperscript{]}}}
\fi

\usepackage{newtxtext,newtxmath}
\usepackage{tikz-cd}
\usepackage{multicol}

\definecolor{isored}{RGB}{215, 38, 61}
\definecolor{okg}{RGB}{27, 127, 75}
\definecolor{nwo}{RGB}{193, 87, 27}
\definecolor{trkc}{RGB}{51, 65, 85}
\definecolor{betac}{RGB}{0, 83, 160}

\author{
    \IEEEauthorblockN{
        \begin{minipage}{0.3\textwidth}
            \centering
            Amro Saleh \\
            \textit{Department of Computer Science} \\
            \textit{Princess Sumaya University for Technology} \\
            Amman 11941, Jordan \\
            amr20229009@std.psut.edu.jo
        \end{minipage}
        \hspace{0.4cm}
        \begin{minipage}{0.3\textwidth}
            \centering
            Loukas Lazos \\
            \textit{School of Electrical, Computer, and Software Engineering} \\
            \textit{The University of Arizona} \\
            Tucson, AZ 85721, USA \\
            llazos@email.arizona.edu
        \end{minipage}
        \hspace{0.4cm}
        \begin{minipage}{0.3\textwidth}
            \centering
            Mohammad J. Abdel-Rahman \\
            \textit{Department of Data Science} \\
            \textit{Princess Sumaya University for Technology} \\
            Amman 11941, Jordan \\
            m.Abdel-Rahman@psut.edu.jo
        \end{minipage}
    }
}

\begin{document}
\title{A Minimally Informed Black-Box Fake-Report Attack on DRL Traffic-Signal Control}
\IEEEoverridecommandlockouts
\IEEEaftertitletext{%
  \begin{center}
    \footnotesize\itshape
    This work has been submitted to the IEEE for possible publication.\\
    Copyright may be transferred without notice, after which this version may no longer be accessible.
  \end{center}%
}
\maketitle

\begin{abstract}
Reinforcement learning (RL) traffic-signal controllers fundamentally change traffic management from a predetermined timing plan to a continuously adapting signal policy \rd{that adapts to}\ra{responsive to} observable traffic conditions. In the absence of sensor instrumentation, the traffic information may originate from vehicle-to-infrastructure (V2I) communication. This new perception layer has been a subject of false data injection attacks aiming to disrupt the flow of traffic. Existing demonstrations typically assume either white-box access to the controller's weights or full observability of the traffic state. We study the impact of a weaker and more realistic adversary who has no access to the RL model and whose traffic observations are limited to her own approach. This black-box, minimally informed attacker trains an imitation surrogate of the controller. She then crafts bounded fake reports, restricted to inflating vehicle counts on her own approach, that transfer to the real controller. We demonstrate that despite the loss of accuracy relative to an attacker \rd{that}\ra{who} can observe the entire state and action space, this limited attacker can significantly increase the average waiting time for other approaches while \rd{reducing waiting on its own approach to nearly zero}\ra{barely changing the waiting time on her own approach, i.e.\ externalizing the added delay onto the other approaches}. \rd{This result shows that the attack requires only the direction of the controller's response, rather than a faithful policy surrogate.} \ra{Because the only realizable move is to inflate one's own reported queue, we find that a zero-knowledge attacker that simply reports the maximum feasible queue is as damaging as the surrogate-based one, while an unstructured random inflation is not: the surrogate is not required to reproduce the degradation, and the exposed vulnerability is the V2I sensing channel itself. The surrogate's role is directional control, selecting a disruptive or a selfish objective at the same budget.} The attack \rd{remains effective under}\ra{persists under} $\pm20\%$ \ra{uniform} demand variation. \rd{Finally, adapting the crafting rules of recent related work to the same setting, we compare their delay impact under a matched budget.}\ra{Finally, on a common per-lane budget we place the attacker on a knowledge ladder against white-box adversaries, and find the black-box surrogate matches a realizable white-box attacker that uses the victim's own gradients; only abandoning realizability raises the damage further.}
\end{abstract}


\section{Introduction}
\begin{figure*}[t]
\centering
\resizebox{\textwidth}{!}{%
\begin{tikzpicture}[
  font=\rmfamily\footnotesize,
  >=Stealth,
  line cap=round,
  line join=round,
  panel/.style={
    draw=black!35, fill=black!2, rounded corners=2pt,
    line width=0.6pt
  },
  compact/.style={
    draw, rounded corners=2pt, line width=0.7pt,
    align=center, text width=25mm, minimum height=24mm,
    inner xsep=1.5mm, inner ysep=1.5mm
  },
  report/.style={
    -{Stealth[length=1.8mm]}, draw=blue!65!black, line width=0.85pt
  },
  attack/.style={
    -{Stealth[length=1.8mm]}, draw=red!70!black, line width=0.85pt
  },
  attackinfo/.style={
    -{Stealth[length=1.8mm]}, draw=red!70!black,
    dashed, line width=0.7pt
  },
  control/.style={
    -{Stealth[length=1.8mm]}, draw=green!50!black, line width=0.85pt
  },
  pics/topcar/.style args={#1/#2}{
    code={
      \begin{scope}[rotate=#2]
        \path[draw=#1!80!black, fill=#1!16, rounded corners=1pt,
              line width=0.55pt]
          (-4.2mm,-2.25mm) rectangle (4.2mm,2.25mm);
        \draw[#1!80!black, line width=0.45pt]
          (-3.5mm,-2.25mm) -- (-3.5mm,2.25mm);
        \draw[#1!80!black, line width=0.45pt]
          (3.5mm,-2.25mm) -- (3.5mm,2.25mm);
        \path[fill=#1!45!black, rounded corners=0.6pt]
          (-1.9mm,-1.55mm) rectangle (1.9mm,1.55mm);
        \draw[white!80, line width=0.35pt]
          (-0.9mm,-1.45mm) -- (-0.9mm,1.45mm);
        \draw[white!80, line width=0.35pt]
          (0.9mm,-1.45mm) -- (0.9mm,1.45mm);
        \fill[black!75] (-2.6mm,-2.45mm) rectangle (-1.3mm,-2.05mm);
        \fill[black!75] (1.3mm,-2.45mm) rectangle (2.6mm,-2.05mm);
        \fill[black!75] (-2.6mm,2.05mm) rectangle (-1.3mm,2.45mm);
        \fill[black!75] (1.3mm,2.05mm) rectangle (2.6mm,2.45mm);
      \end{scope}
    }
  }
]

\node[panel, minimum width=54mm, minimum height=56mm,
      anchor=north west] (environment) at (0,0) {};
\node[panel, minimum width=99mm, minimum height=56mm,
      anchor=north west] (controller)
      at ([xshift=5mm]environment.north east) {};

\coordinate (junction) at (environment.center);

\fill[black!10]
  ($(junction)+(-23mm,-6.5mm)$) rectangle ($(junction)+(23mm,6.5mm)$);
\fill[black!10]
  ($(junction)+(-6.5mm,-23mm)$) rectangle ($(junction)+(6.5mm,23mm)$);

\draw[white, dashed, line width=0.55pt]
  ($(junction)+(-23mm,0)$) -- ($(junction)+(-7mm,0)$);
\draw[white, dashed, line width=0.55pt]
  ($(junction)+(7mm,0)$) -- ($(junction)+(23mm,0)$);
\draw[white, dashed, line width=0.55pt]
  ($(junction)+(0,-23mm)$) -- ($(junction)+(0,-7mm)$);
\draw[white, dashed, line width=0.55pt]
  ($(junction)+(0,7mm)$) -- ($(junction)+(0,23mm)$);
\draw[white, line width=0.9pt]
  ($(junction)+(-7mm,-6mm)$) -- ($(junction)+(-7mm,6mm)$);
\draw[white, line width=0.9pt]
  ($(junction)+(7mm,-6mm)$) -- ($(junction)+(7mm,6mm)$);
\draw[white, line width=0.9pt]
  ($(junction)+(-6mm,-7mm)$) -- ($(junction)+(6mm,-7mm)$);
\draw[white, line width=0.9pt]
  ($(junction)+(-6mm,7mm)$) -- ($(junction)+(6mm,7mm)$);

\node[font=\rmfamily\scriptsize\bfseries, text=black!60]
  at ($(junction)+(0,25mm)$) {N};
\node[font=\rmfamily\scriptsize\bfseries, text=black!60]
  at ($(junction)+(-2mm,-25mm)$) {S};
\node[font=\rmfamily\scriptsize\bfseries, text=black!60]
  at ($(junction)+(25mm,0)$) {E};
\node[font=\rmfamily\scriptsize\bfseries, text=black!60]
  at ($(junction)+(-25mm,0)$) {W};

\coordinate (compromised) at ($(junction)+(-3.3mm,15mm)$);
\pic at (compromised) {topcar={red/90}};
\pic at ($(junction)+(3.3mm,18mm)$) {topcar={blue/90}};
\pic at ($(junction)+(-3.3mm,-17mm)$) {topcar={blue/90}};
\pic at ($(junction)+(3.3mm,-13mm)$) {topcar={blue/90}};
\pic at ($(junction)+(-16mm,-3.3mm)$) {topcar={blue/0}};
\pic at ($(junction)+(-18mm,3.3mm)$) {topcar={blue/0}};
\pic at ($(junction)+(16mm,3.3mm)$) {topcar={blue/180}};
\pic at ($(junction)+(18mm,-3.3mm)$) {topcar={blue/180}};

\node[draw=black!60, fill=black!65, rounded corners=0.8pt,
      minimum width=3.6mm, minimum height=9mm] (signal) at (junction) {};
\fill[red!75!black]    ([yshift=2.4mm]signal.center) circle (0.7mm);
\fill[yellow!70!black] (signal.center) circle (0.7mm);
\fill[green!55!black]  ([yshift=-2.4mm]signal.center) circle (0.7mm);

\node[draw=red!70!black, fill=red!3, rounded corners=2pt,
      line width=0.75pt, align=center, text width=91mm,
      minimum height=15mm, inner xsep=2mm, inner ysep=1.2mm,
      anchor=north] (attacker)
  at ([yshift=-3mm]controller.north)
  {\textbf{\textcolor{red!70!black}{Compromised northbound V2I node}}:
   observes \(o_N,a_t\), trains a local surrogate, and injects
   \(\widetilde{o}_N=o_N+\delta\) using inflate-only PGD.};

\node[compact, draw=blue!65!black, fill=blue!4,
      anchor=south west] (observations)
  at ([xshift=3mm,yshift=4mm]controller.south west)
  {\textbf{\textcolor{blue!65!black}{Observation}}\\
   \(o_S,o_E,o_W,\widetilde{o}_N\)\\
   \(q_\ell,\;w_\ell\)};

\node[compact, draw=violet!70!black, fill=violet!4,
      right=3mm of observations] (agent)
  {\textbf{\textcolor{violet!70!black}{PPO policy}}\\
   \(\pi_\theta(a_t\mid o_t)\)\\
   delay-based reward};

\node[compact, draw=green!50!black, fill=green!4,
      right=3mm of agent] (action)
  {\textbf{\textcolor{green!50!black}{Signal action}}\\
   \(a_t\in\{1,\ldots,8\}\)\\
   hold for \(\Delta t\)};


\draw[report]
  (environment.east |- observations.west)
  -- (observations.west);

\draw[attack]
  (compromised) -- ++(3mm,0)
  |- node[pos=0.67, above, fill=white, inner sep=0.6pt] {\(o_N\)}
  (attacker.west);

\draw[attack]
  (attacker.south -| observations.north)
  -- node[right, inner sep=0.6pt] {\(\widetilde{o}_N\)}
  (observations.north);

\draw[report] (observations.east) -- (agent.west);
\draw[report] (agent.east) -- (action.west);

\draw[attackinfo]
  (action.north)
  -- node[right, inner sep=0.6pt] {\(a_t\)}
  (attacker.south -| action.north);

\draw[control]
  (action.south) -- ++(0,-2.5mm)
  -| node[pos=0.47, above, fill=white, inner sep=0.6pt] {\(a_t\)}
  (signal.south);

\end{tikzpicture}%
}
\vspace{-0.2in}
\caption{PPO-based traffic-signal control under a compromised V2I report. Per-lane queue lengths $q_\ell$ and head-of-queue waiting times $w_\ell$ are assembled from V2I reports and used by the PPO policy to select one of eight conflict-free signal phases for each control interval $\Delta t$. The attacker perturbs only its North-approach report, $\tilde{o}_N=o_N+\delta$, while the remaining approach reports $o_S,o_E,o_W$ are authentic.}
\label{fig:intersection}
\end{figure*}
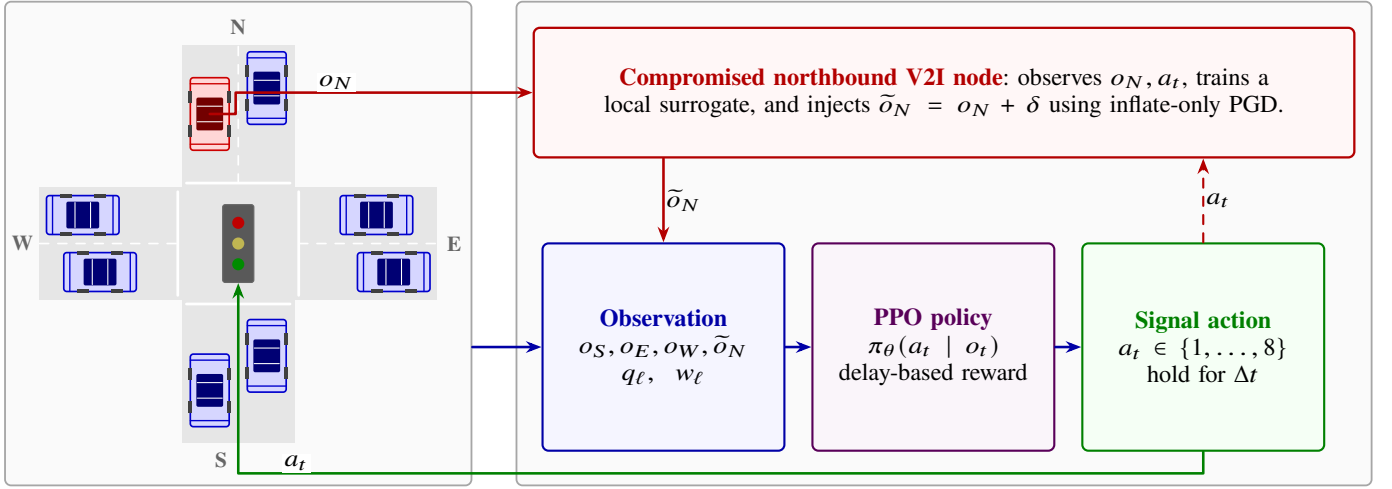

Reinforcement learning (RL)-based traffic signal control provides significant performance advantages compared to predetermined timing or threshold-triggered controllers. An RL controller adapts its phase decisions to the live demand it perceives, by learning optimal policies for maximizing traffic flow and reducing waiting time~\cite{xiao2026survey,wei2018intellilight,WANG2025110440,chen2020mplight}. The framework of an RL-based traffic management for a single intersection is shown in Fig.~\ref{fig:intersection}. An RL controller regulates the traffic phase in all approach directions by observing the per-lane queue lengths and head-of-queue waiting times. Such observations can be collected by physical sensors, when available, or be reported over vehicle-to-infrastructure (V2I) links. The agent selects one signal phase from a fixed conflict-free set and holds it for an interval $\Delta t$ (a short yellow precedes any change), and is rewarded for reducing intersection delay and queue length, learning to maximize the cumulative reward.

For controllers that rely on V2I reports, an adversary who fabricates such reports can manipulate the controller's perceived traffic state and, consequently, the signal phase it selects, without ever accessing the controller. Even if V2I reports are protected cryptographically, the adversary can still use compromised cryptographic credentials to inject false reports. Prior attacks on RL signal control often study the impact of false V2I reports under strong and unrealistic adversary models. These include white-box access to the victim's model weights~\cite{huang2017adv,kos2017delving,xu2023tits}, coordination among many colluding vehicles~\cite{qu2023acm}, or full observation of the traffic state across all approach directions. In this paper, we study a more realistic and low-effort threat in which a single roadside adversary has minimal information about the traffic state of a lane that she can observe and no access to the RL model (black-box attack). For this threat model, we make the following contributions.


\emph{\textbf{Our Contributions:}} 
\begin{itemize}
\item We formalize a realistic black-box, partial-observation threat model against RL-based traffic control. An agent runs a frozen policy $\pi_V$ which is trained observing the per-lane queue lengths and head-of-line waiting time on each approach. The attacker observes only the state of its own approach, has no access to the victim's weights or gradients, and is limited to injecting V2I reports on its own approach. \rd{Moreover, the adversary is aware of the publicly observable controller's actions. By following the phase of her own direction, she can infer the phase of other directions, since the traffic light phases are selected from a conflict-free set.} \ra{Moreover, the adversary observes the publicly actuated phase index $a^V$: signal indications are public, and in our setup the executed phase is directly available to the roadside attacker. We do not assume the global phase is inferred from the attacker's own indication alone.}

\item Exploiting the actuated phase and partial lane traffic information, the adversary trains an imitation surrogate $\hat\pi_V$ of the frozen controller, a separately trained clone (parameters $\phi$) that estimates the victim policy $\pi_V$ from \ra{own-approach} observations only, using supervised cross-entropy (imitation learning)~\cite{papernot2017practical,behzadan2017vulnerability}. She then uses the surrogate policy to select a budget-constrained perturbation by projected gradient ascent on $\hat\pi_V$. The adversary's objective is to steer the victim's decisions in favor of the attacker's preferred phase, projecting the result onto the realizable, budgeted set. No victim weights or gradients are ever used.

\item Applying the attack on traffic collected from a real intersection, we show that this minimally informed black-box attacker is \rd{as damaging as}\ra{comparably damaging to (within ${\sim}1.5$\,s across budgets)} one with full \rd{state}\ra{report} observation, even though its surrogate is a much poorer policy clone. The attacker can selfishly reduce its own waiting time at the expense of the waiting time of other approaches, concentrating the added delay on those approaches. We validate the attack against $\pm20\%$ change in traffic demand to show the vulnerability persists under demand shifts. \rd{Finally, we place the attack on a common vehicle-equivalent budget axis against stronger threat models and compare their delay impact under matched budgets.} \ra{Finally, we situate the attack on a \emph{knowledge ladder} at a common per-lane budget: from a zero-knowledge max-inflation rule (${\sim}17$\,s) and our black-box surrogate (${\sim}15$--$16$\,s) up to a realizable white-box attacker with full gradient access (${\sim}17$\,s) and an unconstrained, non-realizable white-box ceiling (${\sim}56$\,s, over a ${\sim}5$\,s benign baseline). Controller knowledge buys almost nothing once the attack is confined to realizable own-approach inflation, while an unstructured random inflation (${\sim}10$\,s) fails: the danger is the fake-report channel, not attacker sophistication.}
\end{itemize}

\section{Related Work}\label{sec:related}


We group prior attacks on deep reinforcement learning (DRL) signal control into three lines of work by the knowledge and control the adversary must command, the axis that governs whether an attack is realizable in deployment. Observation-perturbation attacks originate with methods that assume
white-box access to the victim's weights or gradients~\cite{huang2017adv,kos2017delving},
later specialized to traffic control by Xu \emph{et~al.}~\cite{xu2023tits} with a Carlini--Wagner
state attack. A second line of work targets the RL-TSC learning signal directly: Ren
\emph{et~al.}\ perturb the reward~\cite{ren2024iotj} or select \emph{critical} states in
multi-agent TSC~\cite{ren2024tii}, and Liu \emph{et~al.}~\cite{liu2025semibb} inject
surrogate-crafted falsified trajectories across a network. A third line assumes the physical capability to \emph{coordinate} many actors: colluding vehicles that jointly shape the reported state~\cite{qu2023acm}, Sybil identities in RL-controlled
traffic~\cite{sybil_rl2022}, or falsified-data injection through the connected-vehicle
channel~\cite{arabi2021csr,huang2020falsified}.

These threats differ most in \emph{how much the adversary must know and control}, as
Table~\ref{tab:relwork} lays out. Along that axis they cluster at the well-informed end: full gradient access, the reward channel, the complete traffic state, or many coordinated nodes.
The opposite, knowledge-restricted end, a single node that observes only its own approach and holds nothing but one budgeted V2I injection channel, is the realistic case and, to our
knowledge, is not studied for traffic-signal control; it is the only configuration in
Table~\ref{tab:relwork} that is simultaneously black-box, single-node, partial-observation,
and realizable. Our adversary sits precisely there.

\begin{table}[t]
\centering
\caption{Related attacks on \ra{learning-based and connected-vehicle} signal control along the knowledge-and-control axis. ``No wts.'': never reads victim weights/gradients; \ra{``Own-appr.''}: needs only the attacker's own approach; ``Realiz.'': perturbs a physically injectable message, not internal state or reward. \ra{Huang \emph{et~al.}: non-RL I-SIG controller.}}
\label{tab:relwork}
\resizebox{\columnwidth}{!}{%
\begin{tabular}{@{}lllccc@{}}
\toprule
Attack (source) & Channel & Nodes & No wts. & \ra{Own-appr.} & Realiz. \\
\midrule
Xu \emph{et~al.}~\cite{xu2023tits}            & state      & one   & $\times$    & $\times$    & $\times$ \\
Ren \emph{et~al.}~\cite{ren2024iotj}          & reward     & multi & \ra{\checkmark}    & $\times$    & $\times$ \\
Ren \emph{et~al.}~\cite{ren2024tii}           & state      & multi & \ra{\checkmark}    & $\times$    & $\times$ \\
Liu \emph{et~al.}~\cite{liu2025semibb}        & trajectory & multi & \checkmark  & $\times$    & \checkmark \\
Qu \emph{et~al.}~\cite{qu2023acm}             & V2I report & multi & \checkmark  & $\times$    & \checkmark \\
Sybil-RL~\cite{sybil_rl2022}                  & V2I report & multi & \checkmark  & $\times$    & \checkmark \\
Huang \emph{et~al.}~\cite{huang2020falsified} & V2I report & multi & \checkmark  & $\times$    & \checkmark \\
\textbf{This work} & \textbf{V2I report} & \textbf{one} & \textbf{\checkmark} & \textbf{\checkmark} & \textbf{\checkmark} \\
\bottomrule
\end{tabular}}
\vspace{-12pt}

\end{table}
It is deliberately the most knowledge-restricted of the set (weak in knowledge, not in access), which is why its effectiveness (Section~\ref{sec:results}) is the surprising result. On the defense side, misbehavior detection~\cite{vanderheijden2019misbehavior}, physics-based attack detection~\cite{giraldo2018physics}, and certified adversarial robustness for deep RL~\cite{lutjens2019certified} target this surface; we operate on the attack side and defer concrete countermeasures to the journal version.

\section{System and Threat Models}\label{sec:threat}
We model three parties: the physical traffic system, the RL victim that controls it, and the adversary that corrupts the victim's reports (Fig.~\ref{fig:intersection}).

\subsection{Physical system}
The controlled plant is a single four-way intersection with four approaches ($N,E,S,W$) and four lanes per approach (Fig.~\ref{fig:intersection}); the two incoming lanes of each approach are signal-controlled and enter the observation (eight controlled lanes). At each decision epoch the signal holds one of eight conflict-free phases for an interval $\Delta t{=}10$\,s\ra{; on a phase change the interval consists of a $3$\,s all-yellow transition followed by $7$\,s of green, and a full $10$\,s of green when the phase is unchanged, so the $3$\,s yellow is taken from, not added to, the interval}. The microscopic traffic state $s\in\mathcal{S}$ is not observed directly; it is reported over the vehicle-to-infrastructure (V2I) channel as a per-lane vector $o=h_V(s)\in\mathbb{R}^{16}$, where the sensing map $h_V$ stacks the queue length $q_\ell$ and head-of-queue waiting time $w_\ell$ of the eight controlled lanes (dimension $16=8\text{ lanes}\times2\text{ features}$), each normalized by $Q_{\mathrm{norm}}{=}40$ vehicles and $W_{\mathrm{norm}}{=}200$\,s. All rewards and reported metrics are computed from the true state $s$, never the reported one.

\subsection{Victim model}
The victim is a Proximal Policy Optimization (PPO) controller acting on the reports $o \in\mathbb{R}^{16}$, from a representative single-intersection, policy-based RL-TSC design~\cite{WANG2025110440,wei2018intellilight} rather than a bespoke target. We cast control as a Markov decision process $(\mathcal{S},\mathcal{A},P,r,\gamma)$ with state space $\mathcal{S}$, action set $\mathcal{A}$, transition kernel $P$, per-step reward $r$, and discount factor $\gamma$. The action $a\in\{\rd{0..7}\ra{1,\dots,8}\}$ selects one of the eight phases, after which the environment advances the real vehicles, $s'\sim P(\cdot\mid s,a)$. \ra{Because the report $o$ omits the current phase and the internal signal timers, control is formally a POMDP; following common practice we treat $o$ as the policy input.} The reward 
\[r_t=\textcolor{revadd}{\frac{\mathit{WT}_{t-\Delta t}-\mathit{WT}_t}{W_{\mathrm{norm}}}+\frac{\mathit{QL}_{t-\Delta t}-\mathit{QL}_t}{Q_{\mathrm{norm}}}}
\]
is the interval reduction in \rd{mean waiting time $\mathit{WT}$ and mean queue length $\mathit{QL}$}\ra{total head-of-queue waiting time $\mathit{WT}$ (seconds) and total queue length $\mathit{QL}$ (vehicles), each normalized by $W_{\mathrm{norm}}{=}200$\,s and $Q_{\mathrm{norm}}{=}40$ so the two terms are dimensionless and commensurate}, and the policy $\pi_V(a| o)$ (weights $\theta$) maximizes the discounted return $\mathbb{E}[\sum_t\gamma^t r_t]$. We attack a frozen policy $\pi_{\textcolor{revadd}{V}}$ at evaluation time.


\subsection{Attacker model}
\emph{Knowledge.} The attacker acts on the victim observation $o$ defined in Section~\ref{sec:threat} above, but sees only its own approach. Its restricted view is the masked observation
\begin{equation}
o_K = m_K \odot o, \qquad m_K \in \{0,1\}^{16},
\label{eq:mask}
\end{equation}
where $\odot$ is the element-wise product and the binary mask $m_K$ keeps the four dimensions of the attacker's own approach, the queue $q_\ell$ and head-of-queue waiting $w_\ell$ of its two lanes, and zeroes the twelve dimensions of the other three approaches. The attacker additionally observes the publicly actuated phase $a^V$ (the victim's executed action $a$), but never the victim weights $\theta$ or gradients. This is a knowledge-restricted adversary, limited to one approach.

 The method by which the attacker turns this \ra{own-approach} view $o_K$ into a fake report, by first training a surrogate $\hat\pi_V$ and then crafting the perturbation, is developed in Section~\ref{sec:attack}.

\emph{Why this model is realistic.} The adversary is a single compromised or colluding
vehicle, or a roadside transmitter, on \emph{one} approach. It naturally sees the V2I
reports and public signal timing of its own approach, so she knows its own
lane state and the phases the controller actuates. She does not have access to the states of other approaches or the controller's weights or gradients. This is strictly weaker than the white-box and full-observation adversaries
assumed in prior work, and it is what a real roadside attacker could actually observe. \ra{Concretely, the controller consumes a per-lane aggregate (queue length and head-of-queue wait) assembled from V2I messages on each approach. We model the attacker as controlling this aggregate for its \emph{own} approach only, realized either as a compromised reporting node that emits the approach's aggregate report (one credential, one inflated report) or as an in-vehicle transmitter that emits several fabricated basic-safety messages (Sybil identities) the aggregator counts as vehicles. The per-lane budget $\varepsilon Q_{\mathrm{norm}}$ bounds the fabricated count in either case; whether it needs one credential or several Sybil identities is an assumption on the reporting pipeline, not a change to the attack. We do not claim that one authenticated vehicle can, by itself, appear as $\varepsilon Q_{\mathrm{norm}}$ distinct authenticated vehicles.}

\emph{Budget and realizability.} The attacker adds a perturbation $\delta$ to its own-approach report, injecting $\tilde o_N=o_N+\delta$ (Fig.~\ref{fig:intersection}), bounded by magnitude $\varepsilon$ ($\varepsilon Q_{\mathrm{norm}}$ fabricated vehicles) and by realizability: perturbations are non-negative and confined to the attacker's own approach. We take inflation as the primary threat because it manufactures demand to seize green, whereas deflating one's own approach is self-defeating; signed, disruption-oriented perturbations are a symmetric extension.

\section{Black-Box Partial-Information Attack}\label{sec:attack}
\emph{Surrogate.} The attacker first builds the surrogate it will attack through (Fig.~\ref{fig:intersection}). From the \ra{own-approach} pairs $(o_K,a^V)$ logged in Section~\ref{sec:threat}, it trains an imitation clone $\hat\pi_V$ of the frozen controller by supervised cross-entropy on the observed phases~\cite{papernot2017practical,behzadan2017vulnerability}; because it sees only its own approach this clone is deliberately weak (imitation accuracy ${\approx}0.79$ versus $0.99$ under full observation), yet Section~\ref{sec:results} shows it suffices.

\begin{figure}[t]
\centering
\begin{tikzpicture}[
  font=\scriptsize,
  >=Stealth,
  trk/.style={->, line width=0.9pt, color=trkc},
  btrk/.style={->, line width=1.1pt, color=betac},
  tk/.style={font=\tiny\bfseries, color=trkc, inner sep=1pt},
  btk/.style={font=\tiny\bfseries, color=betac, inner sep=1pt},
  badge/.style={draw, rounded corners=2pt, font=\tiny\bfseries, inner sep=2pt}
]

  \def\xStart{2.5}
  \def\xEnd{6.75}
  \def\xOp{5.5} 

  \begin{scope}[on background layer]
    \fill[betac!8, rounded corners=3pt] (0.1, 2.05) rectangle (8.4, 2.95);
  \end{scope}

  \draw[trk] (\xStart, 3.5) -- (\xEnd, 3.5);
  \node[anchor=east, align=right] at (2.35, 3.5) {
    \textbf{Vehicle-equivalent}\\[-1pt]
    {\tiny\color{black!60}$\# = \varepsilon Q_{\mathrm{norm}}$ \ra{per lane}}
  };
  \foreach \x/\t in {2.6/2, 3.5/4, 4.5/10, 5.5/20, 6.6/40}{
    \draw[trkc, line width=0.6pt] (\x, 3.42) -- (\x, 3.58);
    \node[tk, above=1pt] at (\x, 3.55) {\t};
  }

  \draw[trk] (\xStart, 2.5) -- (\xEnd, 2.5);
  \node[anchor=east, align=right] at (2.35, 2.5) {
    \textbf{\color{betac}Surrogate (ours, black-box)}\\[-1pt]
    {\tiny\color{black!60}$\ell_\infty$ budget $\varepsilon$}
  };
  \foreach \x/\t in {2.6/{0.05}, 3.5/{0.10}, 4.5/{0.25}, 5.5/{0.50}, 6.6/{1.00}}{
    \draw[trkc, line width=0.6pt] (\x, 2.42) -- (\x, 2.58);
    \node[tk, above=1pt] at (\x, 2.55) {\t};
  }
  \node[badge, draw=okg!80, fill=okg!10, text=okg, anchor=west] at (6.88, 2.5) {\checkmark\ realizable};

  \draw[trk] (\xStart, 1.5) -- (\xEnd, 1.5);
  \node[anchor=east, align=right] at (2.35, 1.5) {
    \textbf{White-box reference}\\[-1pt]
    {\tiny\color{black!60}$\ell_\infty$ budget $\varepsilon$}
  };
  \foreach \x/\t in {2.6/{0.05}, 3.5/{0.10}, 4.5/{0.25}, 5.5/{0.50}, 6.6/{1.00}}{
    \draw[trkc, line width=0.6pt] (\x, 1.42) -- (\x, 1.58);
    \node[tk, above=1pt] at (\x, 1.55) {\t};
  }
  \node[badge, draw=nwo!80, fill=nwo!10, text=nwo, anchor=west] at (6.88, 1.5) {needs weights};

  \draw[btrk] (\xStart, 0.5) -- (\xEnd, 0.5);
  \node[anchor=east, align=right] at (2.35, 0.5) {
    \textbf{\color{betac}Common budget}\\[-1pt]
    {\tiny\color{betac!80!black}$\beta \in [0, 1]$}
  };
  \foreach \x/\t in {2.6/{0.05}, 3.5/{0.10}, 4.5/{0.25}, 5.5/{0.50}, 6.6/{1.00}}{
    \draw[betac, line width=0.8pt] (\x, 0.40) -- (\x, 0.60);
    \node[btk, below=1.5pt] at (\x, 0.42) {\t};
  }

  \draw[isored, dashed, line width=1.1pt] (\xOp, 0.15) -- (\xOp, 3.9);
  \node[badge, draw=isored, fill=white, text=isored, anchor=south, yshift=2pt] at (\xOp, 3.9) {
    Operating point
  };

\end{tikzpicture}
\vspace{-8pt}

\caption{Common vehicle-equivalent budget axis. The $\ell_\infty$ magnitude $\varepsilon$ maps to $\varepsilon Q_{\mathrm{norm}}$ fabricated vehicles \ra{per attacked lane} ($\beta{=}\varepsilon$); the surrogate (realizable) and the white-box reference (needs weights) share this scale. Dashed line: the $\beta{=}0.5$ operating point ($20$/lane). Details in Sec.~\ref{sec:attack}.}
\label{fig:budget-axis}
\vspace{-12pt}
\end{figure}
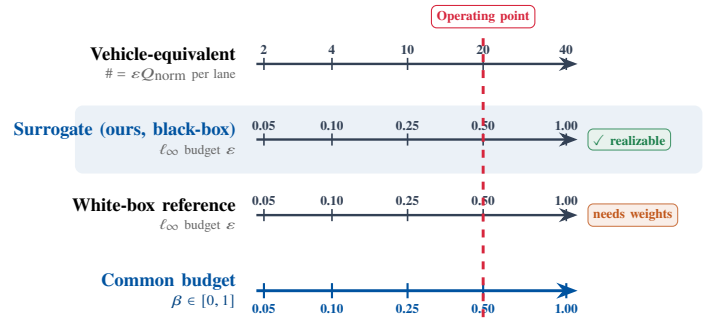

\emph{Attack.} Using the surrogate, the adversary crafts the greedy per-step perturbation
\begin{equation}
\begin{aligned}
\delta^\star &=\arg\max_{\delta\in\mathcal C(\mathcal B)} J(\ra{o_K}+\delta;\,\hat\pi_V),\\
\ra{a^V} &\ra{{}=\arg\max_{a}\,\pi_V(a\mid o+\delta^\star)},
\end{aligned}
\label{eq:attack}
\end{equation}
by projected gradient ascent through $\hat\pi_V$. We take the \emph{disruptive} objective
\begin{equation}
\begin{aligned}
J(\ra{o_K}+\delta;\,\hat\pi_V) &= -\log \hat\pi_V\!\big(a^{\mathrm{cl}}\mid \ra{o_K}+\delta\big),\\
a^{\mathrm{cl}} &=\arg\max_a \hat\pi_V(a\mid \ra{o_K}),
\end{aligned}
\label{eq:objective}
\end{equation}
\ra{The surrogate is always evaluated on the masked view $o_K=m_K\odot o$ (its inaccessible dimensions are zeroed at both training and attack time, so the attacker never uses the other approaches' state); the victim acts on the full report $o+\delta^\star$, whose only altered entries are the attacker's own-approach queues. Because $\delta$ is supported on those queue dimensions, $m_K\odot(o+\delta)=o_K+\delta$. Evaluation is deterministic ($\arg\max$), not sampled.}

i.e.\ the perturbation is pushed to drive the controller off the phase
$a^{\mathrm{cl}}$ it would have taken on the honest report; maximizing $J$ therefore
maximizes the surrogate's probability of a different, and empirically more delaying,
decision. The selfish variant replaces~\eqref{eq:objective} with
$J=\log\!\sum_{a\in\mathcal A_{\mathrm{own}}} \hat\pi_V(a\mid \textcolor{revadd}{o_K}+\delta)$, the log-probability
mass on phases that serve the attacker's own approach; it produces the ``delay for whom''
split of Section~\ref{sec:results}. The victim weights are never used, and the perturbation
transfers to the real controller. The feasible set $\mathcal C(\mathcal B)$
encodes what a roadside adversary can actually inject: a magnitude bound
$\lVert\delta\rVert_\infty\le\varepsilon$ \rd{(equivalently $\varepsilon Q_{\mathrm{norm}}$ fabricated vehicles)} \ra{(an $\ell_\infty$, i.e.\ \emph{per-coordinate}, cap; since the attacked coordinates are the two queue entries of the approach, this permits up to $\varepsilon Q_{\mathrm{norm}}$ fabricated vehicles \emph{on each attacked lane}, hence at most $2\varepsilon Q_{\mathrm{norm}}$ across the approach, not $\varepsilon Q_{\mathrm{norm}}$ in total)} and \emph{realizability}, $\delta\ge0$ restricted to the \rd{four observation dimensions}\ra{two queue dimensions} of the attacker's own approach, since a fabricator can add phantom
vehicles to its own report but cannot erase real ones or edit other approaches. We constrain only the queue dimensions and leave the waiting-time dimensions unperturbed, the more conservative choice; inflating reported waiting time is at least as realizable and would only strengthen the attack. Relaxing realizability to signed, all-dimension perturbations recovers a non-deployable white-box reference. The budget axis of
Fig.~\ref{fig:budget-axis} maps the fabricated vehicles of rule-based attacks and the
$\ell_\infty$ magnitude of gradient attacks to one common scale, so both families are
compared on a common vehicle-equivalent scale.

\begin{algorithm}[t]
\caption{Black-box, partial-observation fake-report attack}
\label{alg:attack}
\begin{algorithmic}[1]
\State \textbf{Input:} frozen victim $\pi_V$; \ra{own-approach} mask $m_K$; budget $\varepsilon$; step size $\alpha$; PGD steps $T_{\mathrm{pgd}}$
\State collect $(o_K^t, a^V_t)$ by observing $\pi_V$ under benign traffic \Comment{own approach only}
\State train surrogate $\hat\pi_V$ to predict $a^V$ from $o_K$ \Comment{imitation, no weights}
\For{each decision with reported state $o$}
  \State $\delta \gets 0$
  \For{$T_{\mathrm{pgd}}$ steps} \Comment{projected gradient ascent on $\hat\pi_V$}
     \State $\delta \gets \delta + \alpha\,\mathrm{sign}\big(\nabla_\delta J(\textcolor{revadd}{o_K}+\delta; \hat\pi_V)\big)$ \Comment{\ra{surrogate sees masked $o_K$}}
     \State $\delta \gets \Pi_{\mathcal C(\mathcal B)}(\delta)$ \Comment{clip to $\varepsilon$ per lane, inflate-only, own queues}
  \EndFor
  \State report $o+\delta$ to $\pi_V$; the true state is unchanged
\EndFor
\end{algorithmic}

\end{algorithm}

Algorithm~\ref{alg:attack} has two stages: an offline stage (lines 2--3) that logs \ra{own-approach}
observation--phase pairs and fits the surrogate $\hat\pi_V$ with no access to victim weights,
gradients, or rewards, and an online stage (lines 4--8) that perturbs each reported state by $T_{\mathrm{pgd}}$ steps of sign-gradient ascent on $\hat\pi_V$ followed by projection onto
$\mathcal C(\mathcal B)$. Only $\hat\pi_V$ is differentiated, so the procedure is black-box with respect to the true controller; the controller acts on the perturbed report, while the environment's true state evolves on the ground truth.
Hyperparameters are listed in Table~\ref{tab:hyper}.

\section{Attack Evaluation}\label{sec:setup}
\subsection{Setup}

We evaluated the attack on a real \texttt{cologne1} intersection from the RESCO
benchmark~\cite{ault2021reinforcement} with a four-approach, eight-phase junction from
the city of Cologne, driven in SUMO from its calibrated network and demand. The victim is
a PPO controller ($Q_{\mathrm{norm}}{=}40$, $\Delta t{=}10$\,s) on a $16$-dimensional
report (queue and head-wait for the eight controlled lanes). Attacks are run over eight
traffic seeds at matched vehicle-equivalent budgets. Our evaluation metrics are: (i) network (global) average waiting time (AWT), \ra{the mean accumulated waiting time over \emph{all} vehicles present in the network;} (ii) the attacker's own-approach AWT, \ra{and} (iii) the other approaches' AWT, \ra{each the mean accumulated waiting time over the vehicles on the controlled lanes of the attacker's approach and of the other three approaches, respectively;} and (iv) the attack activation schedule (fraction of decision steps perturbed). \ra{Metrics (ii)--(iii) cover only vehicles on the eight controlled lanes, a subset of (i), so the global AWT is not a weighted average of the own- and other-approach values.} Using a real benchmark
network keeps the evaluation realistic rather than synthetic. Attacker and surrogate
hyperparameters are given in Table~\ref{tab:hyper}; the surrogate is a small MLP trained by imitation and does not share the victim's PPO network or weights.
\begin{table}[t]
\caption{Attacker, surrogate, and victim hyperparameters.}
\label{tab:hyper}
\centering
\scriptsize
\setlength{\tabcolsep}{3.5pt}
\renewcommand{\arraystretch}{1.1}

\begin{tabularx}{\linewidth}{@{} >{\raggedright\arraybackslash}p{2.3cm} >{\raggedright\arraybackslash}X @{}}
\toprule
\textbf{Component} & \textbf{Hyperparameters \& Settings} \\
\midrule

Surrogate $\hat\pi_V$ & 
MLP $16\!\to\!128\!\to\!128\!\to\!8$ (ReLU); $300$ epochs, Adam $10^{-3}$, class-weighted CE; $2$ benign imitation episodes (held-out split) \\
\addlinespace[2pt]

PGD Attack & 
$T_{\mathrm{pgd}}{=}10$, $\alpha{=}1.5\varepsilon/T_{\mathrm{pgd}}$, $\varepsilon \in \{0.1, 0.25, 0.5\}$ ($4, 10, 20$ veh/lane at $Q_{\mathrm{norm}}{=}40$) \\
\addlinespace[2pt]

Victim Policy & 
PPO actor--critic (SB3 \texttt{MlpPolicy}, $[64,64]$ tanh); frozen, zero weight/gradient access \\
\addlinespace[2pt]

Victim Training & 
$10^5$ steps ($980$ updates, final checkpoint); Adam lr $3{\times}10^{-4}$, $\gamma{=}0.99$, GAE $\lambda{=}0.95$, $n_{\mathrm{steps}}{=}1024$, batch $256$, clip $0.2$, ent $0.01$, epochs $10$, $v_f$ $0.5$ \\

\bottomrule
\end{tabularx}
\vspace{-8pt}
\end{table}

\subsection{Results}\label{sec:results}
\subsubsection{Benign Validation} 

Before any attack, the trained victim is stable: its benign global AWT is $5.3$\,s across the eight seeds ($95\%$ confidence interval (CI) $[5.0,5.5]$\,s)\rd{, confirming a well-trained controller,}\ra{. On the same scenario a fixed-time controller yields $38.9$\,s, so the PPO victim is roughly $7.3\times$ better and is a credible rather than brittle baseline;} so the degradation reported below is attributable to the attack and not to a weak baseline. \ra{To confirm the vulnerability is structural rather than specific to one checkpoint, we repeated the black-box partial attack on five independently trained PPO victims (distinct seeds, identical training recipe): benign global AWT $5.4\pm0.2$\,s rises to $11.9\pm1.3$\,s at the ten-vehicle budget, with every victim degraded (attacked AWT $10.0$--$13.3$\,s) and the delay consistently externalized onto the other approaches (others $11.3$\,s vs.\ own $1.9$\,s). The exposure is thus a property of the fake-report channel, not of one checkpoint; the victim analyzed in detail below is among the more susceptible.}

\subsubsection{Imitation under Partial Observation}

Trained on its own approach only, the surrogate imitates the controller with accuracy $0.79$, versus $0.99$ when trained on the full report. A central finding, developed next, is that this limited clone is enough: the attack
needs the perturbation's \emph{direction} on the attacker's own approach, not a faithful
copy of the whole policy.

\subsubsection{Partial Observation is \ra{Nearly} Enough}

The \ra{own-approach} attacker is \rd{matches the full-observation attacker at every budget}\ra{comparable to the full-observation attacker across the tested budgets} (Table~\ref{tab:partial}, Fig.~\ref{fig:dose}(a)): at ten vehicles it
raises global AWT to \rd{$15.3\pm0.4$\,s (vs. $16.5\pm0.2$\,s}\ra{$15.4\pm0.4$\,s (vs. $16.7\pm1.0$\,s} under complete observation), and at twenty vehicles to \rd{$27.4\pm1.6$\,s (vs. $26.1\pm0.5$\,s)}\ra{$27.2\pm1.0$\,s (vs. $26.7\pm2.1$\,s)} from a benign $5.3$\,s (mean $\pm$ s.d.\ over $8$ seeds). \rd{At ten vehicles the two knowledge levels are practically equivalent within a pre-registered $2$\,s margin, established by a paired two one-sided tests procedure (TOST, a statistical equivalence test). At twenty vehicles the two means differ by only $1.3$\,s ($27.4$ vs.\ $26.1$\,s), which is within the same $2$\,s margin; we report this descriptively and do not claim statistical equivalence at this budget, because a paired TOST here requires the per-seed partial/complete results (a non-significant difference and overlapping s.d.\ ranges are not equivalence evidence).}\ra{A paired two one-sided tests procedure (TOST; $16$ seeds; pre-specified $2$\,s margin) does \emph{not} certify equivalence at either budget (ten vehicles: mean difference $-1.5$\,s, $90\%$ CI $[-2.3,-0.7]$; twenty vehicles: $+1.2$\,s, CI $[-0.1,+2.4]$): the own-approach attacker is slightly weaker at the low budget and slightly stronger at the high budget, with means within ${\sim}1.5$\,s throughout. We therefore describe the two settings as \emph{comparable}, not statistically equivalent.} Either way, restricting the adversary to its \ra{own approach} does not protect the controller\ra{: a surrogate that is a far poorer clone ($0.79$ vs.\ $0.99$) comes within ${\sim}1.5$\,s of the full-observation attack, and at the higher budget is marginally stronger}.

\begin{table}[t]
\centering
\caption{\ra{Own-approach} (partial) vs.\ full-observation (complete) black-box attack on \texttt{cologne1} (mean $\pm$ s.d., $8$ seeds; benign $5.3$\,s). ``Acc'': imitation accuracy; AWT in seconds.}
\label{tab:partial}
\begin{tabular}{@{}lccc@{}}
\toprule
Knowledge (obs.\ access) & Acc & AWT @10 veh & AWT @20 veh \\
\midrule
Complete (full report)   & 0.99 & \ra{$16.7\pm1.0$} & \ra{$26.7\pm2.1$} \\
\textbf{Partial (\ra{own approach} only)} & 0.79 & \ra{$\mathbf{15.4\pm0.4}$} & \ra{$\mathbf{27.2\pm1.0}$} \\
\bottomrule
\end{tabular}
\vspace{-8pt}
\end{table}

\subsubsection{Robustness to Demand}

To check that the vulnerability is not an artifact of the calibrated demand, we replay the same victim under SUMO's global demand scaled to $0.8\times$ and $1.2\times$ (Table~\ref{tab:demand}). The attack degrades the controller across the whole range\rd{, and the effect is stronger than the raw numbers suggest: the attacked AWT is almost demand-invariant ($\approx15.4$\,s at ten vehicles and $\approx28$\,s at twenty across all three demand levels), while the benign AWT climbs from $2.9$ to $9.3$\,s. In other words, the own-lane attacker collapses the intersection to the same congested ceiling regardless of how much real traffic is present, so the ratio over benign is largest (over $10\times$ at twenty vehicles) precisely when the road is quiet}\ra{: the degradation \emph{relative} to benign is largest when the road is quiet ($7.1\times$ at $0.8\times$ vs.\ $3.2\times$ at $1.2\times$ demand); the attacked AWT is not fully demand-invariant, but a large degradation persists at every level (Table~\ref{tab:demand})}. The delay-for-whom split holds at every demand level (other approaches \rd{$15$--$16$}\ra{$10$--$17$}\,s vs. the attacker's own \rd{$1$--$2$}\ra{$1$--$5$}\,s
at ten vehicles).
\begin{table}[t]
\centering
\caption{Demand sensitivity (uniform $\pm20\%$ scaling; mean $\pm$ s.d., $8$ seeds; AWT in seconds). Last column: ratio of attacked AWT@20 to benign at the same demand (from unrounded means).}
\label{tab:demand}
\begin{tabular}{@{}lcccc@{}}
\toprule
Demand & Benign & AWT @10 veh & AWT @20 veh & AWT@20 / benign \\
\midrule
$0.8\times$ & \ra{$2.9$} & \ra{$10.8\pm0.5$} & \ra{$20.9\pm0.5$} & \ra{$7.1\times$} \\
$1.0\times$ & $5.3$ & \ra{$15.4\pm0.4$} & \ra{$27.2\pm1.0$} & \ra{$5.1\times$} \\
$1.2\times$ & \ra{$8.3$} & \ra{$17.4\pm2.8$} & \ra{$26.3\pm1.3$} & \ra{$3.2\times$} \\
\bottomrule
\end{tabular}
\vspace{-15pt}
\end{table}

\subsubsection{Delay for whom?}

Global AWT understates the attack's character. Splitting the waiting time by approach at the ten-vehicle budget (Fig.~\ref{fig:perapproach}(b)), the attacker's
own approach falls to $1.7$\,s\rd{, below the benign $5.3$\,s,}\ra{ (against a global benign mean of $5.3$\,s; a per-approach benign reference is not yet measured, see caption)}, while the three other
approaches climb to $14.8$\,s. The mechanism is direct: by inflating its own reported queue
the adversary persuades the controller to over-serve its approach, so the disruption is
paid almost entirely by the other directions. \ra{Instrumenting the controller confirms this mechanism and rules out a phase-chattering artifact: under attack the attacker's green share rises from $0.46$ (benign) to $0.66$ ($10$-veh) and $0.72$ ($20$-veh), while the phase-switch rate \emph{falls} ($0.82\!\to\!0.54\!\to\!0.42$) and the yellow-time fraction drops ($0.25\!\to\!0.16\!\to\!0.13$). The added delay is caused by over-serving the compromised approach, not by extra switching or lost yellow time.} This asymmetry has two consequences. It shows
the attack can be \emph{selfish} (win green for oneself) as easily as \emph{disruptive}
(raise network delay), depending only on which approach is compromised; and \rd{it means a network-average monitor would see a modest global rise while one corridor is being systematically starved, making the attack harder to notice than its global number suggests.}\ra{because the added delay concentrates on the uncompromised approaches, the per-approach picture is more severe for those directions than the global figure implies. The global AWT itself rises sharply ($5.3\!\to\!15.3$\,s, $+189\%$), so the attack is not hidden at the network level; the split shows the burden is borne by the other approaches rather than shared.}

\begin{figure*}[t]
\centering
\begin{minipage}[b]{0.49\textwidth}
\centering
\begin{tikzpicture}
\begin{axis}[
  width=\linewidth,
  height=4.6cm,
  xlabel={Attack budget (vehicle-equivalent)},
  ylabel={Global AWT (s)},
  xmin=0, xmax=21,
  ymin=0, ymax=30,
  xtick={0, 4, 10, 20},
  ytick={0, 5, 10, 15, 20, 25, 30},
  tick label style={font=\footnotesize},
  label style={font=\small},
  axis lines=left,
  grid=major,
  grid style={line width=0.3pt, gray!20},
  legend cell align={left},
  legend style={
    font=\scriptsize,
    at={(0.03, 0.96)},
    anchor=north west,
    draw=black!20,
    fill=white,
    fill opacity=0.9,
    text opacity=1,
    inner sep=3pt,
    row sep=1pt
  },
  every axis plot/.append style={line width=1.2pt, mark size=2.0pt}
]

\addplot[cgray, dashed, line width=0.9pt, domain=0:21, samples=2, mark=none] {5.3};
\addlegendentry{Benign baseline (5.3\,s)}

\addplot[cred, mark=*, mark options={fill=cred},
  error bars/.cd, y dir=both, y explicit]
  coordinates {(4, 9.35) +- (0,2.24) (10, 15.35) +- (0,0.37) (20, 27.17) +- (0,1.04)};
\addlegendentry{Partial (own approach, acc 0.79)}

\addplot[cblue, densely dashed, mark=square*, mark options={fill=cblue},
  error bars/.cd, y dir=both, y explicit]
  coordinates {(4, 8.74) +- (0,0.11) (10, 16.71) +- (0,1.03) (20, 26.70) +- (0,2.13)};
\addlegendentry{Complete (full report, acc 0.99)}

\end{axis}
\end{tikzpicture}\\[-2pt]
{\footnotesize (a) Dose--response}
\end{minipage}\hfill
\begin{minipage}[b]{0.49\textwidth}
\centering
\begin{tikzpicture}
\begin{axis}[
  width=\linewidth,
  height=4.6cm,
  ybar,
  bar width=11pt,
  enlarge x limits=0.5,
  ylabel={AWT (s)},
  ymin=0, ymax=18,
  ytick={0, 3, 6, 9, 12, 15, 18},
  symbolic x coords={Benign, Attacked},
  xtick=data,
  tick label style={font=\footnotesize},
  label style={font=\small},
  axis lines*=left,
  ymajorgrids=true,
  xmajorgrids=false,
  grid style={line width=0.3pt, gray!20},
  legend cell align={left},
  legend style={
    font=\scriptsize,
    at={(0.03, 0.96)},
    anchor=north west,
    draw=black!20,
    fill=white,
    fill opacity=0.9,
    text opacity=1,
    inner sep=3pt,
    row sep=1pt
  },
  nodes near coords,
  nodes near coords style={
    font=\tiny\bfseries,
    /pgf/number format/fixed,
    /pgf/number format/precision=1,
    yshift=1pt
  }
]

\addplot[fill=cred!75, draw=cred!90!black, line width=0.5pt] 
  coordinates {(Benign, 2.2) (Attacked, 1.7)};
\addlegendentry{Attacker's own approach}

\addplot[fill=cblue!65, draw=cblue!90!black, line width=0.5pt] 
  coordinates {(Benign, 3.2) (Attacked, 14.8)};
\addlegendentry{Other approaches}

\end{axis}
\end{tikzpicture}\\[-2pt]
{\footnotesize (b) Delay for whom}
\end{minipage}
\vspace{-4pt}
\caption{\emph{(a)} Dose--response on \texttt{cologne1}: the \ra{own-approach} (partial) attacker matches the full-observation attacker across the budget range, despite seeing only its own approach. \emph{(b)} Delay for whom (partial attack, $10$-vehicle budget; \ra{benign per-approach over $16$ seeds}): \ra{the attacker's own approach barely changes ($2.2\!\to\!1.7$\,s) while the other approaches are starved ($3.2\!\to\!14.8$\,s), externalizing the added delay rather than reducing the attacker's own wait, an asymmetry the global AWT hides.}}
\label{fig:dose}
\label{fig:perapproach}
\vspace{-8pt}
\end{figure*}

\subsection{Benchmarking}\label{sec:compare}
\ra{We read the results as a \emph{knowledge ladder}: how much must an attacker know before the fake-report channel becomes dangerous? At a matched per-lane budget and a ${\sim}5$\,s benign baseline, the rungs are an unconstrained white-box ceiling that ignores realizability (${\sim}56$\,s), a realizable white-box attacker with full victim gradients (${\sim}17$\,s), our black-box own-approach surrogate (${\sim}15$--$16$\,s), a zero-knowledge max-inflation rule (${\sim}17$\,s), and an unstructured random inflation that fails (${\sim}10$\,s). The ordering shows that realizability, not knowledge of the controller, is what bounds the damage: every realizable attacker, from full-gradient to zero-knowledge, lands within a few seconds of the others.}
The closest attacks (Table~\ref{tab:relwork}) differ from ours mainly in perturbation
channel and in requiring more access~\cite{ren2024iotj,ren2024tii,xu2023tits,qu2023acm,liu2025semibb};
several are multi-agent or multi-intersection but reduce to our single-agent setting. To compare
crafting rules fairly, we reimplement two representative mechanisms in our setup, holding
intersection, victim, budget, seeds, and \ra{own-approach} observation fixed and swapping only the
rule: a Jacobian saliency rule after Liu \emph{et~al.}\ (\textsc{jsma}) and a critical-step rule
after Ren \emph{et~al.}, in the partial-information black-box setting (Table~\ref{tab:compare}). Our
projected-gradient surrogate and the JSMA saliency rule are \ra{comparable in delay (means within $0.5$\,s; Table~\ref{tab:compare})}: the advantage of the PGD surrogate is not its AWT magnitude but its simpler crafting, needing no saliency-map construction and working purely from the \ra{own-approach} gradient direction. \ra{The most informative rows are the two zero-knowledge baselines: a \emph{max-inflate} rule (no surrogate, gradients, or action observation) that reports the maximum feasible queue every step is at least as damaging as the surrogate, whereas \emph{random} inflation is much weaker (Table~\ref{tab:compare}). On this inflate-only surface the gradient perturbation merely saturates the two queue coordinates; what the surrogate adds is directional control (disruptive vs.\ selfish) and a lower activation rate. The severe vulnerability is a property of the fake-report channel, not of crafting sophistication.} The critical-step rule fires on only about \rd{$9\%$}\ra{$4\%$} of
steps: on this well-trained \texttt{cologne1} clone almost no decision falls inside its logit margin, so it barely perturbs the controller. Its schedule is therefore knowledge-sensitive and would need per-scenario threshold re-tuning, which we report honestly rather than tune away. \ra{Because PGD, JSMA, max-inflate, and random attack $0.76$--$1.0$ of steps while critical-step attacks only ${\sim}0.04$, Table~\ref{tab:compare} is a mechanism comparison at a common per-lane budget, not a schedule-matched efficacy ranking.} \rd{The realizability-relaxed white-box attack of Fig.~\ref{fig:budget-axis} is omitted from Table~\ref{tab:compare}: it requires the victim's weights and is a non-deployable reference, not a black-box competitor.}\ra{A separate $10$-seed white-box comparison at ten vehicles confirms the ladder: our surrogate ($16.4$\,s) is statistically indistinguishable from a realizable, inflate-only white-box PGD using the victim's gradients ($17.0$\,s), so weight access buys almost nothing once the attack is confined to realizable own-approach inflation; only the unconstrained, non-realizable ceiling ($56.4$\,s vs.\ benign $5.2$\,s) does more.}

\begin{table}[t]
\centering
\caption{Related crafting mechanisms on \texttt{cologne1} (partial-information black-box; mean, $8$ seeds; benign $5.3$\,s). ``Sched.'': fraction of steps attacked; AWT global (s); ``own/oth'': own vs.\ other AWT @10 veh.}
\label{tab:compare}
\resizebox{\columnwidth}{!}{%
\begin{tabular}{@{}llcccc@{}}
\toprule
Method (source) & Know. & Sched. & AWT@10 & AWT@20 & own/oth@10 \\
\midrule
PGD surrogate (ours)               & black-box & \rd{every}\ra{$0.77$}    & \ra{$15.4\pm0.4$} & \ra{$27.2\pm1.0$} & \ra{1.7 / 14.8} \\
JSMA saliency~\cite{liu2025semibb} & black-box & \rd{every}\ra{$0.76$}    & \ra{$15.2\pm0.3$} & \ra{$27.7\pm2.9$} & \ra{1.7 / 14.7} \\
Critical-step~\cite{ren2024tii}    & black-box & \rd{$9\%$}\ra{$0.04$}    & \ra{$5.2\pm0.1$}  & \ra{$5.3\pm0.2$}  & \ra{2.0 / 3.3}  \\
\ra{Max-inflate (rule)}~           & \ra{none (0-knowl.)} & \ra{$1.00$} & \ra{$17.5\pm2.2$} & \ra{$28.4\pm3.6$} & \ra{2.1 / 18.3} \\
\ra{Random inflation}~             & \ra{none (0-knowl.)} & \ra{$1.00$} & \ra{$9.8\pm0.3$}  & \ra{$16.2\pm2.0$} & \ra{1.1 / 9.2} \\
\bottomrule
\end{tabular}}
\vspace{-8pt}
\vspace{-8pt}

\end{table}

The per-approach split of Section~\ref{sec:results} (Fig.~\ref{fig:perapproach}(b)) holds for
every crafting rule: each loads far more delay on the other approaches than on the
attacker's own, so the selfish-versus-disruptive duality is a property of the fake-report
channel, not of any one perturbation rule.

\section{Discussion and Limitations}\label{sec:disc}
{\bf Why Own-\rd{Lane}\ra{Approach} Information is Enough.} The attack does not need a faithful clone of the
controller\rd{; it needs only the sign of the perturbation on its own approach, i.e.\ which way to bend its reported queue so the controller keeps serving it. That direction is recoverable from own-lane observations alone, which is why the partial-information surrogate (accuracy $0.79$) matches the full-observation one (accuracy $0.99$) in delay.}\ra{: the partial-information surrogate (accuracy $0.79$) matches the full-observation one (accuracy $0.99$) in delay, and, more strikingly, a zero-knowledge attacker that always reports the maximum feasible own-approach queue matches the surrogate (Table~\ref{tab:compare}). Because the only realizable move is to inflate one's own queue, the perturbation direction is fixed a priori (upward) and need not be learned; only unstructured inflation fails, since it spreads rather than concentrates the budget. The surrogate stays useful for choosing which objective to pursue (disruptive vs.\ selfish) and for attacking fewer steps, but it is not what makes the attack effective.} This is the practical danger: the weakest attacker one can build, a single compromised vehicle on one approach\ra{, with no model of the controller at all,} is already as effective as a far better-informed one.

{\bf Threat Surface and Realism.} The perturbation lives entirely in the V2I report and is
constrained to be realizable: inflate-only, on the attacker's own approach, within a vehicle-equivalent budget. It never touches the victim's weights, gradients, or reward, and the true state and all metrics evolve on the ground truth. This keeps the model consistent with a connected-vehicle deployment in which the sensing channel, not the controller, is the compromised component.

{\bf Limitations.} We re-implemented each competitor's crafting rule inside our setting rather than its full end-to-end system, so the comparison isolates the perturbation mechanism at matched budget and is not a claim about their complete pipelines. Our JSMA is a single-step saliency variant, and the critical-step rule's schedule depends on the clone's logit margin, which on this well-trained victim leaves few decisive steps; both would need per-scenario re-tuning to be pushed harder. Results are for a single real intersection (\texttt{cologne1}); we vary demand by $\pm20\%$ (Table~\ref{tab:demand}) and the attack holds, but multi-intersection coordination and a wider range of networks are left to the future work, together with the defenses. \ra{All results use a single trained PPO checkpoint; confirming the vulnerability across independently trained victim seeds is left to the extended version. The phase-switching, green-share, and yellow-time diagnostics that separate over-serving from phase chattering are reported in Section~\ref{sec:results}.}

\section{Conclusion and Future Work}
A black-box adversary that sees only its own \rd{lane}\ra{approach} and never accesses the victim's weights degrades a PPO traffic-signal controller \rd{as much as}\ra{comparably to (within ${\sim}1.5$\,s across budgets)} a full-observation attacker. On the real \texttt{cologne1} intersection it inflates only its own reported queue, yet holds the attacked delay near a fixed ceiling across $\pm20\%$ \ra{uniform} demand variation and matches a \rd{white-box reference on a common vehicle-equivalent budget}\ra{matched, realizable (inflate-only) white-box attacker (within seed noise at the ten-vehicle budget), while remaining far below an unconstrained white-box ceiling that abandons realizability}, even though its imitation surrogate stays a deliberately poor clone of the victim\ra{, and even a zero-knowledge max-inflation attacker is as damaging}. The harm is also targeted: the compromised approach is \rd{served below its benign wait}\ra{barely affected} while the others are starved\rd{, a split that a network-average monitor would miss}\ra{; the global waiting time rises sharply, so the attack is visible in aggregate, yet the per-approach concentration, borne by the uncompromised approaches, is hidden within that global figure}. Because \rd{the perturbation only needs the right direction on its own approach}\ra{inflating one's own reported queue is by itself enough}, restricting the attacker's view does not defend the controller; the exposed surface is the connected-vehicle reporting channel, not the policy itself, and a single node with minimal knowledge and one realizable injection point is enough to exploit it.

Future work moves toward defenses and scale. The reports themselves can be screened for plausibility, for instance cross-approach consistency and physical limits on how fast a queue can grow, as a cheap first filter, while adversarial or robust training hardens the policy; both should be measured against an adaptive attacker that reacts to the defense rather than a fixed one. At the network level, the same fabricated reports can propagate between signals, so multi-intersection and multi-agent control may either contain or compound the disruption. Finally, detection that infers the attacker's intent from its behavioral signature, rather than the raw report values, is a natural step toward countermeasures that hold up against a knowledgeable adversary.

\bibliographystyle{IEEEtran}
\bibliography{references}

@article{qu2023acm,
  author  = {Qu, Ao and Tang, Yihong and Ma, Wei},
  title   = {Adversarial Attacks on Deep Reinforcement Learning-Based Traffic Signal Control Systems with Colluding Vehicles},
  journal = {ACM Trans. Intell. Syst. Technol.},
  volume  = {14}, number = {6}, articleno = {113}, year = {2023}, doi = {10.1145/3625236}}

@article{xu2023tits,
  author  = {Xu, Dongwei and Li, Chengbin and Wang, Da and Gao, Guangyan},
  title   = {Robustness Analysis of Discrete State-Based Reinforcement Learning Models in Traffic Signal Control},
  journal = {IEEE Trans. Intell. Transp. Syst.},
  volume  = {24}, number = {2}, pages = {1710--1723}, year = {2023}, doi = {10.1109/TITS.2022.3221107}}

@article{ren2024tii,
  author  = {Ren, Yan and Zhang, Heng and Du, Linkang and Zhang, Zhikun and Zhang, Jian and Li, Hongran},
  title   = {Stealthy Black-Box Attack with Dynamic Threshold Against {MARL}-Based Traffic Signal Control System},
  journal = {IEEE Trans. Ind. Informat.},
  volume  = {20}, number = {10}, pages = {12021--12031}, year = {2024}, doi = {10.1109/TII.2024.3413356}}

@article{ren2024iotj,
  author  = {Ren, Yan and Zhang, Heng and Cao, Xianghui and Yang, Chaoqun and Zhang, Jian and Li, Hongran},
  title   = {Promoting or Hindering: Stealthy Black-Box Attacks Against {DRL}-Based Traffic Signal Control},
  journal = {IEEE Internet Things J.},
  volume  = {11}, number = {4}, pages = {5816--5825}, year = {2024}, doi = {10.1109/JIOT.2023.3308260}}

@misc{sybil_rl2022,
  author  = {Irfan, Muhammad Sami and Rahman, Mizanur and Atkison, Travis and Dasgupta, Sagar and Hainen, Alexander},
  title   = {Reinforcement Learning Based Cyberattack Model for Adaptive Traffic Signal Controller in Connected Transportation Systems},
  howpublished = {arXiv:2211.01845}, year = {2022}, doi = {10.48550/arXiv.2211.01845}}

@misc{huang2020falsified,
  author  = {Huang, Shihong E. and Wong, Wai and Feng, Yiheng and Chen, Qi Alfred and Mao, Z. Morley and Liu, Henry X.},
  title   = {Impact Evaluation of Falsified Data Attacks on Connected Vehicle Based Traffic Signal Control Systems},
  howpublished = {Proc. NDSS Workshop Automot. Auton. Veh. Secur. (AutoSec)}, year = {2021}}

@inproceedings{arabi2021csr,
  author  = {Seifollahpour Arabi, Najmeh and Halabi, Talal and Zulkernine, Mohammad},
  title   = {Reinforcement Learning-Driven Attack on Road Traffic Signal Controllers},
  booktitle = {Proc. IEEE Int. Conf. Cyber Secur. Resilience (CSR)}, pages = {218--225}, year = {2021}, doi = {10.1109/CSR51186.2021.9527951}}

@misc{huang2017adv,
  author  = {Huang, Sandy and Papernot, Nicolas and Goodfellow, Ian and Duan, Yan and Abbeel, Pieter},
  title   = {Adversarial Attacks on Neural Network Policies},
  howpublished = {Proc. Int. Conf. Learn. Represent. (ICLR), Workshop Track}, year = {2017}}

@misc{kos2017delving,
  author  = {Kos, Jernej and Song, Dawn},
  title   = {Delving into Adversarial Attacks on Deep Policies},
  howpublished = {Proc. Int. Conf. Learn. Represent. (ICLR), Workshop Track}, year = {2017}}

@misc{lutjens2019certified,
  author  = {L{\"u}tjens, Bj{\"o}rn and Everett, Michael and How, Jonathan P.},
  title   = {Certified Adversarial Robustness for Deep Reinforcement Learning},
  howpublished = {Proc. Conf. Robot Learn. (CoRL), PMLR vol.~100, pp.~1328--1337}, year = {2019}}

@article{giraldo2018physics,
  author  = {Giraldo, Jairo and Urbina, David and Cardenas, Alvaro and Valente, Junia and
             Faisal, Mustafa and Ruths, Justin and Tippenhauer, Nils Ole and
             Sandberg, Henrik and Candell, Richard},
  title   = {A Survey of Physics-Based Attack Detection in Cyber-Physical Systems},
  journal = {ACM Comput. Surv.},
  volume  = {51}, number = {4}, articleno = {76}, year = {2018}, doi = {10.1145/3203245}
}

@article{vanderheijden2019misbehavior,
  author  = {van der Heijden, Rens W. and Dietzel, Stefan and Leinm{\"u}ller, Tim and Kargl, Frank},
  title   = {Survey on Misbehavior Detection in Cooperative Intelligent Transportation Systems},
  journal = {IEEE Commun. Surveys Tuts.},
  volume  = {21}, number = {1}, pages = {779--811}, year = {2019}, doi = {10.1109/COMST.2018.2873088}
}

@inproceedings{papernot2017practical,
  author    = {Papernot, Nicolas and McDaniel, Patrick and Goodfellow, Ian and
               Jha, Somesh and Celik, Z. Berkay and Swami, Ananthram},
  title     = {Practical Black-Box Attacks against Machine Learning},
  booktitle = {Proc. ACM Asia Conf. Comput. Commun. Secur. (ASIA CCS)},
  pages     = {506--519}, year = {2017}, doi = {10.1145/3052973.3053009}
}

@inproceedings{behzadan2017vulnerability,
  author    = {Behzadan, Vahid and Munir, Arslan},
  title     = {Vulnerability of Deep Reinforcement Learning to Policy Induction Attacks},
  booktitle = {Proc. Int. Conf. Mach. Learn. Data Mining (MLDM)},
  series    = {LNCS}, volume = {10358}, pages = {262--275}, year = {2017},
  doi       = {10.1007/978-3-319-62416-7_19}
}

@article{WANG2025110440,
  author  = {Wang, Lijuan and Zhang, Guoshan and Yang, Qiaoli and Han, Tianyang},
  title   = {An Adaptive Traffic Signal Control Scheme with Proximal Policy Optimization Based on Deep Reinforcement Learning for a Single Intersection},
  journal = {Eng. Appl. Artif. Intell.},
  volume  = {149}, pages = {110440}, year = {2025}, doi = {10.1016/j.engappai.2025.110440}
}

@inproceedings{wei2018intellilight,
  author    = {Wei, Hua and Zheng, Guanjie and Yao, Huaxiu and Li, Zhenhui},
  title     = {{IntelliLight}: A Reinforcement Learning Approach for Intelligent Traffic Light Control},
  booktitle = {Proc. ACM SIGKDD Int. Conf. Knowl. Discovery Data Mining (KDD)},
  pages     = {2496--2505}, year = {2018}, doi = {10.1145/3219819.3220096}
}

@inproceedings{chen2020mplight,
  author    = {Chen, Chacha and Wei, Hua and Xu, Nan and Zheng, Guanjie and Yang, Ming and
               Xiong, Yuanhao and Xu, Kai and Li, Zhenhui},
  title     = {Toward a Thousand Lights: Decentralized Deep Reinforcement Learning for Large-Scale Traffic Signal Control},
  booktitle = {Proc. AAAI Conf. Artif. Intell.},
  volume    = {34}, number = {4}, pages = {3414--3421}, year = {2020}, doi = {10.1609/aaai.v34i04.5744}
}

@inproceedings{ault2021reinforcement,
  author    = {Ault, James and Sharon, Guni},
  title     = {Reinforcement Learning Benchmarks for Traffic Signal Control},
  booktitle = {Proc. Adv. Neural Inf. Process. Syst. (NeurIPS), Datasets Benchmarks Track},
  year      = {2021}
}

@inproceedings{liu2025semibb,
  author    = {Liu, Xin and Yu, Chunhui and Su, Zhaocheng and Ma, Weidong},
  title     = {A Semi-Black-Box Approach for Falsified Trajectory Injection Attack Against Network Signal Control},
  booktitle = {Proc. IEEE 28th Int. Conf. Intell. Transp. Syst. (ITSC)},
  year      = {2025},
  doi       = {10.1109/ITSC60802.2025.11423694}
}

@article{xiao2026survey,
  author  = {Xiao, Hang and Li, Huale and Wang, Zhaobin and Yang, Zhen and Qi, Shuhan and Zhang, Jiajia and Cai, DingZhong and Yin, JiaQi},
  title   = {Intelligent Traffic Signal Control Based on Reinforcement Learning: A Survey},
  journal = {Artif. Intell. Rev.},
  volume  = {59},
  number  = {5},
  articleno = {128},
  year    = {2026},
  doi     = {10.1007/s10462-026-11530-9}
}
\end{document}